\documentclass[aps,prb,reprint]{revtex4-1}
\pdfoutput=1

\usepackage{amsmath}
\usepackage{amsfonts}
\usepackage{amssymb}
\usepackage{braket}
\usepackage{relsize}
\usepackage{hyperref}

\begin{document}


\title{Addendum to ``Super-GCA from $\mathcal{N}=(2,2)$ super-Virasoro": Super-GCA connection with tensionless strings}

\author{Ipsita Mandal}
\email{imandal@perimeterinstitute.ca}
\affiliation{Perimeter Institute for Theoretical Physics, Waterloo, Ontario N2L 2Y5, Canada}

\begin{abstract}
In this addendum, we consider the connection between certain $2d$ super-GCA, obtained from the parametric contractions of $2d$ SCFTs, which can describe the constraint algebra of null spinning strings.
\end{abstract}

\maketitle


\section{Introduction}
\label{intro}

The bosonic and supersymmetric versions of Galiliean Conformal Algebra (GCA), obtained by a parametric contraction of the corresponding ``parent" relativistic conformal and superconformal groups, have been studied in great details in the recent literature ~\cite{gca,ba09,ba10,bagchi11,bagchi13,bagchi15,hagen,barnich,oblak,luk,luk2,duval,ayan,mart,rouhani,ali,sgca4d,ma10,ma16,aizawa0,gao,aizawa1,masterov12,masterov15,sakaguchi}. The initial aim of the group contraction was to obtain algebras exhibiting non-relativistic conformal symmetry. However, in later studies~\cite{bagchi13,bagchi15}, it was pointed out that an opposite limit of the group contraction of the holomorphic and anti-holomorphic copies of the $2d$ bosonic CFT, dubbed as the ``Ultra-Relativistic" limit, yields the constraint algebra of the tensionless limit of bosonic strings.

The tensionless limit of bosonic and fermionic strings (as well as their quantization) has been a well-studied topic~\cite{schild,isberg93,saltidis96,gamboa89,gamboa90,neto90,lindstrom91,saltidis95,bagchi13,bagchi15}. The tension of a string is in a sense equivalent to the mass of a particle, and hence a tensionless/null string in string theory corresponds to a massless particle in particle physics. The tension going to zero can be interpreted as the string consisting of a collection of massless particles, but subject to constraints as the string is a continuous object even in the tensionless case and these pieces are still connected to each other. This limit has been thought to be useful in studying the high energy behaviour of string theory.

In this addendum, we will consider the connection between certain $2d$ super-GCA, obtained from the parametric contractions of $2d$ SCFTs, which can describe the contraint algebra of null spinning strings. A spinning string is one which has manifest worldsheet supersymmetry. Focussing on the Neveu-Schwarz sectors admitting the use the superspace coordinate formalism, we will consider primary superfields and their representation theory. All these can be obtained from scaling the original superspace coordinates in conformity with the scaling of the generators of the parent SCFT.

\section{Null spinning string with $N$ supersymmetries}
\label{nsusy}

The null spinning string with $N$ supersymmetries, and thus having a global $O(N)$ symmetry, has been considered in~\cite{lindstrom91,saltidis95}. The non-zero (anti-)commutators are given by:
\begin{equation}
\begin{split}
\label{sgca}
& [L_m, L_n] =  (m-n)\, L_{m+n} + ( c_3   m^3 + c_1   m) \, \delta_{m+n,0}\, ,\\
& [L_m, M_n]  =  (m-n) \,M_{m+n} \,,\\
&  \left\{ G_r^j , G_s^l \right \}  = \delta_{j   l} \, M_{ r + s} \,, \quad
[L_m, G_r^j ] =  \left ( \frac{m} {2} - r \right ) \, G^j_{ m+r } \, ,
\end{split}
\end{equation}
with $j,l=(1,2,\ldots,N)$. We will show that these can be obtained by taking the appropriate group contraction of the $\mathcal{N}=(\tilde N , N-\tilde N )$ (where $\tilde N =(0,1,\ldots,N)$ ) super-Virasoro algebra. Here we will consider the null superstring constraint algebras obtained from the $\mathcal{N}=(1,0)$ and $\mathcal{N}=(1,1)$ parent SCFTs. The bosonic subalgebra is identical to the $2d$ GCA obtained earlier with the non-relativistic limit in mind~\cite{ba10}. Hence, we will refer the above supersymmetric algebra as ``super-GCA'' (SGCA) in this Letter.


\section{Tensionless limit of $ (1,1)$ SCFT }
\label{algebra1}

In this section, we show how we can obtain the null spinning string with $N=2$ supersymmetries from the group contraction of $\mathcal{N} = (1,1) $ SCFT.
The holomorphic $\mathcal{N} = 1$ super-Virasoro algebra~\cite{qiu84,qiu85,qiu86,berhadsky,goddard,sotkov} is given by:
\begin{equation}
\begin{split}
\label{sva}
& [\mathcal{L}_m, \mathcal{L}_n] =  (m-n) \mathcal{L}_{m+n} 
+ \frac{c}{8} \, m \, (m^2 - 1) \, \delta_{m+n,0} \, ,\\
& [\mathcal{L}_m, \mathcal{G}_r ]  = \left(\frac{m}{2} - r\right) \mathcal{G}_{m+r}  \,, \\ 
& \left\{ \mathcal{G}_r, \mathcal{G}_s \right\} = 2 \, \mathcal{L}_{r+s} 
 + \frac{c}{2}\left(r^2 - \frac{1}{4}\right)\delta_{r+s,0} \, ,
\end{split}
\end{equation}
where $m, n \in \mathbb{Z} $ and $ r, s \in \mathbb{Z} + \lambda $ (such that $\lambda = 0$ in the Ramond sector and $ \lambda  = \frac{1}{2} $ in the Neveu-Schwarz sector). The corresponding anti-holomorphic sector will be denoted by barred generators: $ \bar{\mathcal{L}}_n $ and $ \bar{\mathcal{G}}_r $. For the Neveu-Schwarz sector, these generators can be expressed in terms of the superspace coordinates as:
\begin{equation}
\begin{split}
\label{svagen}
& \mathcal{L}_n 
\equiv - z^{n+1} \partial_z - \frac{n+1} {2} z^n \theta \, \partial_\theta \,,\quad
 \mathcal{G}_r 
\equiv  z^{ r+ \frac{1} {2} } \left ( \partial_{\theta } - \theta  \partial_z \right ) \,,\\
& \bar{\mathcal{L} }_n 
\equiv  - \bar{z}^{n+1} \partial_{ \bar{z}} - \frac{n+1} {2} \bar{z}^n \bar{\theta} \, \partial_{ \bar{\theta}}
 \,,\quad
 \bar{ \mathcal{G}}_r 
\equiv  \bar{z}^{ r+ \frac{1} {2} } \left ( \partial_{\bar{ \theta } } - \bar{ \theta}  \partial_{\bar{z}} \right ) \,.
\end{split}
\end{equation}
Considering the Euclidean worldsheet coordinates on a cylinder, which describe closed strings, we take $z = e^{i  \omega }$ and $\bar{ z } = e^{i \bar{\omega} }$, such that $ \omega = \tau + \sigma $ and $\bar{\omega}= \tau-\sigma$. The fermionic coordinates on the cylinder are $\tilde \theta = \frac{\theta }{\sqrt {z}}$ and $\bar{\tilde \theta }= \frac{\bar{\theta} }{\sqrt {\bar{z} }}$. Hence, we have to use $\partial_z = - e^{i \omega}\left ( i \, \partial_\omega  + \frac{\tilde \theta } {2} \, \partial_{\tilde \theta } \right )$, $\partial_\theta = e^{ i \omega/2}   \partial_{\tilde \theta} $ and so on.

The generators for a null spinning string with two supersymmetries~\cite{gamboa89,gamboa90,neto90} obey the algebra in Eq.~(\ref{sgca}), with $N=2$. It is very easy to see that it can be obtained from
a group contraction of the $\mathcal{N} =(1,1)$ SCFT as:
\begin{equation}
\begin{split}
\label{sgcadef}
L_n &= \lim_{\epsilon \to 0} \,
( \mathcal{L}_n  - \bar{\mathcal{L}}_{-n} )\, , 
 \quad M_n = \lim_{\epsilon \to 0} \, \epsilon \, ( \mathcal{L}_n  + \bar{\mathcal{L}}_{-n} )\, ,\\
G_r^1 &= \lim_{\epsilon \to 0} \, \sqrt{ \epsilon} \,  \mathcal{G}_r \, ,  
\quad 
G_r^2 = \lim_{\epsilon \to 0} \, \sqrt{   \epsilon} \,  \bar{\mathcal{G}}_{-r} \, .\\
\end{split}
\end{equation}
In terms of the superspace coordinates, this group contraction corresponds to the scaling:
\begin{equation}
\begin{split}
\label{sgcacoord}
& \tau \rightarrow \epsilon \, \tau \,, \quad
\, \sigma \rightarrow \sigma \,, 
\\ & 
\alpha_1 = \frac{ \tilde \theta } {\sqrt {2} }  \rightarrow \sqrt{\epsilon} \, \alpha_1 \,,
\quad
 \alpha_2 = \frac{ \bar{\tilde \theta} } {\sqrt {2} }\rightarrow \sqrt{\epsilon} \,  \alpha_2 \,.
\end{split}
\end{equation}
This can be interpreted as the``Ultra-Relativistic" limit describing a tensionless closed string (see~\cite{isberg93,saltidis96,bagchi13,ba14,bagchi15} for the bosonic case). The group generators on the cylinder can then be expressed as:
\begin{equation}
\begin{split}
\label{sgcagen}
& L_n \equiv 
e^{ i n \sigma }
\Big \lbrace \
i \, \partial_\sigma - n \left ( \tau \,  \partial_{\tau}
+\frac{ \alpha_j } {2}  \, \partial_{\alpha_j }
\right )
\Big \rbrace \,,\\ & 
 M_n \equiv  i \, e^{ i n \sigma }  \partial_\tau \,, 
\quad 
 G_{r}^j \equiv
\frac{ e^{ i r  \sigma}}  { \sqrt {2}  }
\left(   
 \partial_{\alpha_j } + i\, \alpha_j  \, \partial_\tau 
\right)  \, .
\end{split}
\end{equation}
The same generators, when written for the plane, take the form:
\begin{equation}
\begin{split}
\label{sgcagenp}
& L_n \equiv 
- x^{ n+1 } \partial_x
- (n+1) \, x^n \left( t \, \partial_t
+ \frac{\alpha_j^p }{2} \, \partial_{\alpha_j^p } \right )
 \,,\\ & 
 M_n \equiv  - x^{ n+1 }  \partial_t \,, 
\quad 
 G_{r}^j \equiv
\frac{ x^{ r +\frac{1} {2} }}  { \sqrt {2}  }
\left(   
 \partial_{\alpha_j^p } - \alpha_j^p  \, \partial_t 
\right)  \, ,
\end{split}
\end{equation}
where the two sets of coordinates are related by
\begin{equation}
\begin{split}
\label{map}
& x= e^{i \sigma} \,,
\quad t = i \, \tau \, e^{i \sigma}\,,
\quad \alpha_j^p =  \alpha_j  \, e^{ i \sigma/2}.
\end{split}
\end{equation}

Let us also derive the expressions for the stress-energy tensors ($T_L, T_M$) and their fermionic superpartners ($T_F^j $) from the super-Virasoro stress-energy tensors on a cylinder, which is again relevant for a closed superstring. Though the bosonic part follows from the derivation in~\cite{bagchi13}, we show the complete analysis for the sake of completeness. The holomorphic CFT stress-energy tensor $(T_B)$ has and its fermionic superpartner $( T_F) $ have weights $(2,0)$ and $(3/2,0)$ respectively. Similarly, their anti-holomorphic counterparts ($ \bar{T}_B$ and $\bar{T}_F$) have weights $(0,2)$ and $(0,3/2)$ respectively. Hence, ignoring the central charges, we have the relations:
\begin{equation}
\begin{split}
\label{cyl1}
& T_B^{ \text{cyl} } (\omega) = z^2 \, T_B^{\text{plane} } (z)
= \sum_n  \mathcal{L}_n e^{-i n \omega}
 \,, \nonumber \\
& T_F^{ \text{cyl} } (\omega) = z^{\frac{3}{2} } \, T_F^{\text{plane} } (z) 
= \sum_r  \mathcal{G}_r e^{-i r \omega}\,, \nonumber \\
& \bar{T}_B^{ \text{cyl} } (\bar{\omega} ) = \bar{z}^2 \, \bar{T}_B^{\text{plane} } (\bar{z} )
= \sum_n  \bar{ \mathcal{L}  }_n e^{-i n \bar{\omega} }
 \,, \nonumber \\ 
& \bar{T}_F^{ \text{cyl} } (\bar{\omega}) = \bar{ z }^{\frac{3}{2} } \, \bar{T}_F^{\text{plane} } (\bar{ z }) 
\sum_r  \bar{ \mathcal{G}  }_r e^{-i r  \bar{\omega} }   \,,
\end{split}
\end{equation}
where the superscript ``cyl" stands for cylinder.
Using Eqs.~(\ref{sgcadef}) and (\ref{sgcacoord}), the null spinning string stress-energy tensors and their fermion superpartners are given by:
\begin{equation}
\begin{split}
\label{cyl2}
T_L &= \lim_{\epsilon \to 0} \,
( T_B^{ \text{cyl} }   - \bar{T}_B^{ \text{cyl} }  ) 
= \sum_n \left( L_n - i n \tau M_n
\right ) e^{-i n \sigma }
 , \\ \nonumber
 T_M &= \lim_{\epsilon \to 0} \, \epsilon \, ( T_B^{ \text{cyl} }  + \bar{T}_B^{ \text{cyl} }  )
= \sum_n  M_n e^{-i n \sigma } ,
 \\ \nonumber 
T_F^j &= \lim_{\epsilon \to 0} \, \sqrt{ \epsilon} \,  T_F^{\text{cyl}} 
 = \sum_r  G_r^1 e^{-i r \sigma } ,  \\ \nonumber
T_F^2 & = \lim_{\epsilon \to 0} \, \sqrt{ \epsilon} \,  \bar{T}_F^{\text{cyl}} 
 = \sum_r  G_r^2 e^{-i r \sigma } . 
\end{split}
\end{equation}

\subsection{Representation Theory for super-GCA with two supersymmetries}
After finding an algebra which generates the symmetry of a system, it is worthwhile to develop its representation theory which will tell us about the states that can exist in the model. This subsection is devoted towards finding the representation theory of the SGCAs under study.

The vacuum state $|0\rangle$ should be defined such that the above operators have vanishing vaccum expectation values:
\begin{equation}
\begin{split}
& \langle 0|\, T_L \, | 0 \rangle = \langle 0| \, T_M \, | 0 \rangle =\langle 0| \,  T_F^j \, | 0 \rangle = 0  \\ \nonumber
& \Rightarrow  L_n | 0 \rangle =M_n | 0 \rangle =0 \, ( \text{ for }  n\geq -1)  \\ \nonumber
& \text{and } G_r^{j} | 0 \rangle =0 \, ( \text{ for } r\geq -\frac{1} {2} ) \,,
\end{split}
\end{equation}
so that it is invariant under the global part of the super-GCA, i.e.\ it is annihilated by the ten generators $L_{\pm 1}, L_0,
M_{\pm1}, M_0, G^1_{\pm \frac{1}{2} }, G^2_{\pm \frac{1}{2} }$. The physical states are the ones constructed by acting with $L_{-n}, M_{-n}, G^j_{-r}$ on $ | 0 \rangle $, where $n, r > 0$. The BRST charge ($\Omega$) for the null spinning string was constructed in~\cite{gamboa89,gamboa90,saltidis95} and physical states satisfy $ \Omega \,| \text{phys} \rangle = 0$.

This entire construction now allows to define representations with states $ | h_L, h_M \rangle$ having definite eigenvalues of $L_0$ and $M_0$~\cite{ba09,ba10,bagchi13,ba14}:
\begin{equation}
L_0   | h_L, h_M \rangle = h_L  | h_L, h_M \rangle , \quad M_0   | h_L, h_M \rangle = h_M  | h_L, h_M \rangle .
\end{equation}
We note that
\begin{equation}
h_L=  \lim_{\epsilon \to 0} \, (h  - \bar{h}) \,, \quad 
h_M = \lim_{\epsilon \to 0} \, \epsilon   (h  +  \bar{h}) \,,
\end{equation}
such that ($h, \bar{h}$) are the conformal weights of the corresponding state in the parent CFT (when such a parent state exists).
In particular, demanding that the dimension of the states be bounded from
below, we define primary states $| p \rangle \equiv |h_L, h_M \rangle_p$ as obeying :
\begin{equation}
\begin{split}
  L_n | p \rangle =M_n | p \rangle =G_r^{j} | p \rangle =0 \, ( \text{ for } n, r > 0 ) \,.
\end{split}
\end{equation}
As usual, we can now build an irreducible representation of this SGCA by acting on $|p \rangle $ with $L_{-n}, M_{-n}, G^j_{-r}$ (for $n, r > 0$), discarding the descendants with zero norm (null states).

\section{Tensionless limit of $ (1,0) \text{ or } (0,1) $ SCFT and connection with super-$\text{BMS}_3$ algebra}
\label{algebra2}

A null spinning string with one supersymmetry can be obtained from the group contraction of either
$ \mathcal{N} = (1,0) \text{ or } (0,1) $ SCFT in a way analogous to the discussion in Sec.~\ref{algebra1}.
This same SGCA also coincides with the minimal supersymmetric extension of the $\text{BMS}_3$ algebra~\cite{bms31,bms32,bms33}, which is the algebra of surface charges of the three-dimensional asymptotically flat $\mathcal{N} = 1$ supergravity.

\vspace{1 cm}

\textbf{Note Added}
\\This algebra was found by us while addressing a referee's query during the reviewing of the letter ``Super-GCA from $ \mathcal{N} =(2,2)$ super-Virasoro" \cite{ips16}. However, because this algebra relates to tensionless strings, which is inapplicable to the algebra discussed in the letter, it was not included therein. Recently, we have found two pre-prints \cite{casali,bagchi16} that address this same algebra and therefore believe it important and timely to suggest this note as an addendum to our existing work given we had independently arrived at it earlier.

\clearpage


\bibliography{tensionless}


\end{document}